# Unifying relation for quantum systems driven out of equilibrium

Hiroshi Matsuoka

Department of Physics, Illinois State University, Normal, Illinois, 61790-4560, U.S.A.

## Abstract

We extend a classical relation by Crooks to quantum systems and show that it unifies the Crooks transient fluctuation theorem and the Kawasaki non-linear response relation, which leads to the standard linear response theory. We also show that the Green-Kubo formula for a system in a steady state driven by a *time-independent* driving force is always expressed in terms of the *symmetrized* correlation function for a quantity induced by the driving force and that a generalized entropy for systems in steady heat conduction states satisfies extensions of the Clausius and the Gibbs relations *exactly*.

## I. Introduction

Since 1993, a number of exact relations including the fluctuation theorems [1 – 4] and the Jarzynski equality [5] have been derived first for classical systems driven out of equilibrium and later for quantum systems [6]. For classical stochastic systems, Crooks [7] has derived a relation, which we will call the Crooks-Kawasaki relation (CKR), from what is equivalent to the detailed fluctuation theorem [4] and shown that its special cases include the Crooks transient fluctuation theorem [2], the integral fluctuation theorem [3], from which the Jarzynski equality follows, and the Kawasaki non-linear response relation [3 (a), 8].

For quantum systems, the Crooks transient fluctuation theorem [9], the integral fluctuation theorem [3 (b), 9 (a), 9 (c) – (e), 10], the Jarzynski equality [9 (a), 9 (b), 9 (d), 9 (e), 9 (h), 9 (i), 9 (k), 10 (b), 11, 12 (b)], and the fluctuation theorem for currents of energy and particles in a system driven to a steady state by contact with heat and particle reservoirs [12 (a), 12 (c)] have been derived through various routes all involving the principle of microreversibility [6]. For an isolated system driven by a *time-dependent* external field, Bochkov and Kuzovlev [12 (e)] have derived a general relation whose special case can be regarded as a quantum extension of the Kawasaki non-linear response relation, which Andrew and Gaspard [12 (c)] have also derived and employed to recover the standard linear response theory [12 (d)], where the Green-Kubo formula for a linear response coefficient [12 (e)] is expressed in terms of the *canonical* correlation function for a quantity induced by the external field.

## II. Summary of our results

For quantum systems, using the Crooks transient fluctuation theorem, we can derive the integral fluctuation theorem, the Jarzynski equality, and the fluctuation theorem for the currents, but *not* the Kawasaki non-linear response relation, from which we cannot derive the Crooks transient fluctuation theorem either. A question is then whether there exists a unifying relation whose special cases include both the Crooks transient fluctuation theorem and the Kawasaki non-linear response relation.

In this article, we will show that a quantum extension of the CKR, Eq. (5.1) below, which follows directly from the principle of microreversibility, is indeed such a unifying relation whose special cases include both the Crooks transient fluctuation theorem and the Kawasaki non-linear response relation. Furthermore, we will show that the fluctuation theorem for currents of heat and particles also follows directly from a special case of the quantum CKR so that we can greatly simplify its proof. The quantum CKR therefore not only makes the derivations of all these results simple and transparent but also provides a succinct synthesis of all the results that follow from the Crooks transient fluctuation theorem and the standard linear response theory derived from the Kawasaki non-linear response relation. Although not discussed in this article, the CKR can be also extended to continuously monitored quantum systems [9 (j), 11 (e)]. As the results in this article can be readily modified for classical systems, the CKR is in fact a unifying relation for both classical and quantum systems driven out of equilibrium.

In addition, for quantum systems in steady states, we will present two new results derived from the CKR. First, we will show that for a system driven to a steady state by a *time-independent* driving force, we can always prove the fluctuation theorem for a quantity induced by the driving force so that the Green-Kubo formula for its linear response coefficient is always expressed in terms of the *symmetrized* correlation function of the induced quantity. For a system driven to a steady heat conduction state by a temperature difference between two heat reservoirs, we will show that the fluctuation theorem for the heat current allows us to express the Green-Kubo formula for its thermal conductivity in terms of the *symmetrized* correlation function for the heat current density operator. Furthermore, as an example for systems with steady inputs of work, we will consider a fluid driven to a steady shear flow state by a constant velocity of a plate moving above the fluid. Using the CKR, we will prove the fluctuation theorem for the shear stress on the fluid and obtain the Green-Kubo formula for its shear viscosity in terms of the *symmetrized* correlation function for its shear stress operator.

Secondly, for systems in steady heat conduction states, we will show that a generalized entropy used by Tasaki [9 (b)] for quantum systems driven out of equilibrium is in fact a state variable for the steady states and *exactly* satisfies extensions of the Clausius and the Gibbs relations, which suggests that this generalized entropy may be a more natural extension of equilibrium entropy to steady heat conduction states than a different entropy recently shown to satisfy extensions of the Clausius and the Gibbs relations up to the second order in the heat current [13].

## III. The total Hamiltonian and its eigenstates

### A. The total Hamiltonian

As the CKR is a general universal relation for both isolated systems and systems attached to heat and/or particle reservoirs, we will use an adjective "total" when our discussion applies to the both types of systems. By switching on a perturbation term in its total Hamiltonian, we can drive a system out of its initial equilibrium state and let it go through a non-equilibrium forward process over a time interval $[0,\tau]$. During this time interval, the total system evolves according to its total Hamiltonian $H(t)$, which may consists of three terms: $H^{\mathrm{s}}(t)$ for the system subject to a time-dependent external field, $H^{\mathrm{r}}$ for the reservoirs attached to the system, and $H^{\mathrm{int}}$ for a coupling between the system and the reservoirs. Generally, we will not assume $H^{\mathrm{int}}$ to be weak compared to $H^{\mathrm{s}}(t)+H^{\mathrm{r}}$.

### B. The initial eigenstate

Before the initial time $t=0$, we set $H^{\mathrm{int}}=0$ so that the system is detached from the reservoirs. Just before $t=0$, through a measurement of both the energy and the number of particles in the system as well as those in the reservoirs, we find the total system to be in an initial state $|i\rangle$, which is selected by an equilibrium density matrix $\tilde{\rho}_{\mathrm{in}}$ from the eigenstates of $H^{\rm s}(0)+H^{\rm r}$ and the particle number operators, $\tilde{N}^{\rm s}$ and $\tilde{N}^{\rm r}$, for the system and the reservoirs. Depending on how we prepare the initial equilibrium state for the system and the reservoirs, we choose an appropriate equilibrium density matrix $\tilde{\rho}_{\rm in}$.

For example, consider a system attached to two reservoirs, A and B, which will be discussed in Sec.IX. For simplicity, let us assume the system Hamiltonian $H^{\rm s}$ as well as the reservoir Hamiltonians, $H^{\rm A}$ and $H^{\rm B}$, to be all time-independent. Before $t=0$, we keep the system in equilibrium with a heat and particle reservoir whose inverse temperature and chemical potential are set at $\beta^{\rm s}$ and $\mu^{\rm s}$ while keeping the *k*-th reservoir ($k=\mathrm{A,B}$) in equilibrium with a heat and particle reservoir whose inverse temperature and chemical potential are set at $\beta^k$ and $\mu^k$.

Just before $t=0$, through a measurement of both the energy and the number of particles in the system, we find the system to be in an initial state $|i^{\rm s}\rangle$, which is selected from the eigenstates of the system Hamiltonian $H^{\rm s}$ and the system particle number operator $\tilde{N}^{\rm s}$ by an initial density matrix $\tilde{\rho}_{\rm in}^{\rm s}$ corresponding to a grand canonical ensemble specified by $\beta^{\rm s}$ and $\mu^{\rm s}$. Also through a measurement of both the energy and the number of particles in the reservoirs, we find the *k*-th reservoir to be in an initial state $|i^k\rangle$, which is selected from the eigenstates of the reservoir Hamiltonian $H^k$ and the reservoir particle number operator $\tilde{N}^k$ by an initial density matrix $\tilde{\rho}_{\rm in}^k$ corresponding to a grand canonical ensemble specified by $\beta^k$ and $\mu^k$. The initial eigenstate $|i\rangle$ for the total system is then $|i\rangle=|i^{\rm s}\rangle|i^{\rm A}\rangle|i^{\rm B}\rangle$ and the initial density matrix $\tilde{\rho}_{\rm in}$ for the total system is $\tilde{\rho}_{\rm in}=\tilde{\rho}_{\rm in}^{\rm s}\tilde{\rho}_{\rm in}^{\rm A}\tilde{\rho}_{\rm in}^{\rm B}$.

### C. The final eigenstate

After the final time $t=\tau$, we also set $H^{\rm int}=0$ so that the system is detached from the reservoirs. Just after $t=\tau$, through a measurement of both the energy and the number of particles in the system and those in the reservoirs, we find the total system to be in a final state $|f\rangle$, which is an eigenstate of $H^{\rm s}(\tau)+H^{\rm r}$, $\tilde{N}^{\rm s}$, and $\tilde{N}^{\rm r}$.

For the system attached to two reservoirs, A and B, discussed above, just after $t=\tau$, through a measurement of both the energy and the number of particles in the system and those in the reservoirs, we find the system to be in an eigenstate $|f^{\mathrm{s}}\rangle$ of the system Hamiltonian $H^{\mathrm{s}}$ and the system particle number operator $\tilde{N}^{\mathrm{s}}$ while we find the $k$-th reservoir to be in an eigenstate $|f^{k}\rangle$ of the reservoir Hamiltonian $H^{k}$ and the reservoir particle number operator $\tilde{N}^{k}$. The final eigenstate $|f\rangle$ for the total system is then $|f\rangle = |f^{\mathrm{s}}\rangle |f^{\mathrm{A}}\rangle |f^{\mathrm{B}}\rangle$.

## IV. Time-reversed process and the principle of microreversibility

During $[0,\tau]$, the state $|\Psi(t)\rangle$ of the total system evolves according to the Schrödinger equation with $H(t)$ so that its final state is related to its initial state by $|\Psi(\tau)\rangle = U|\Psi(0)\rangle$, where $U$ is the time evolution operator at $t = \tau$. Since the time reversal operator $\Theta$ satisfies $i\Theta = -\Theta i$ and $\Theta\Theta^{\dagger} = \Theta^{\dagger}\Theta = I$, the time-reversed state, $|\Psi_{\mathrm{r}}(t)\rangle \equiv \Theta|\Psi(\tau - t)\rangle$, evolves according to the Schrödinger equation with the time-reversed Hamiltonian,

$$ {}^{\Theta}H(\tau - t) \equiv \Theta H(\tau - t)\Theta^{\dagger}, \tag{4.1}$$

and the final state of the time-reversed backward process is related to its initial state by $|\Psi_{\mathrm{r}}(\tau)\rangle = {}^{\Theta}U|\Psi_{\mathrm{r}}(0)\rangle$, where ${}^{\Theta}U$ is the time evolution operator at the end of the backward process. We do not assume $H(t)$ to be invariant with respect to time reversal so that $H(t)$ does not need to be equal to $\Theta H(t)\Theta^{\dagger}$. For any $|\Psi(0)\rangle$ we then find

$$ \Theta|\Psi(0)\rangle = |\Psi_{\mathrm{r}}(\tau)\rangle = {}^{\Theta}U|\Psi_{\mathrm{r}}(0)\rangle = {}^{\Theta}U\Theta|\Psi(\tau)\rangle = {}^{\Theta}U\Theta U|\Psi(0)\rangle \tag{4.2}$$

so that

$$ {}^{\Theta}U = \Theta U^{\dagger}\Theta^{\dagger}, \tag{4.3}$$

which is called the principle of microreversibility and is the basis for the CKR.

Using this equation, we can show that the transition probability for the forward process from an initial eigenstate $|i\rangle$ to its final eigenstate $|f\rangle$ is equal to the transition probability for the backward process from $|{}^{\Theta}f\rangle \equiv \Theta|f\rangle$ to $|{}^{\Theta}i\rangle \equiv \Theta|i\rangle$:

$$|\langle f|U|i\rangle|^2 = |\langle {}^{\Theta}i|{}^{\Theta}U|{}^{\Theta}f\rangle|^2, \tag{4.4}$$

which follows from

$$\langle {}^{\Theta}i|{}^{\Theta}U|{}^{\Theta}f\rangle = \left(\Theta|i\rangle,\ \Theta U^{\dagger}|f\rangle\right) = \left(U^{\dagger}|f\rangle,\ |i\rangle\right) = \langle f|U|i\rangle, \tag{4.5}$$

where $\Theta$ is anti-unitary so that for any pair of states, $|\alpha\rangle$ and $|\alpha'\rangle$, $\Theta$ satisfies $\left(\Theta|\alpha'\rangle,\ \Theta|\alpha\rangle\right) = \left(|\alpha\rangle,\ |\alpha'\rangle\right)$, where $\left(|\alpha\rangle,\ |\alpha'\rangle\right)$ is the inner product between $|\alpha\rangle$ and $|\alpha'\rangle$.

The following property of $\Theta$ will be also useful. If $|n\rangle$ is an eigenstate of an observable $A$ with a real eigenvalue $a(n)$ so that $A|n\rangle = a(n)|n\rangle$, then $|{}^{\Theta}n\rangle \equiv \Theta|n\rangle$ is an eigenstate of ${}^{\Theta}A \equiv \Theta A \Theta^{\dagger}$ so that ${}^{\Theta}A|{}^{\Theta}n\rangle = {}^{\Theta}a({}^{\Theta}n)|{}^{\Theta}n\rangle$ and ${}^{\Theta}a({}^{\Theta}n) = a(n)$ because

$${}^{\Theta}A|{}^{\Theta}n\rangle = \Theta A|n\rangle = a(n)\Theta|n\rangle = a(n)|{}^{\Theta}n\rangle. \tag{4.6}$$

## V. The Crooks-Kawasaki relation and the total entropy production

### A. The Crooks-Kawasaki relation

The Crooks-Kawasaki relation (CKR) relates the forward process average involving a quantity $C(i,f)$ that depends on the initial and final eigenstates, $|i\rangle$ and $|f\rangle$, of the total system to its average over the backward process from $|{}^{\Theta}f\rangle$ to $|{}^{\Theta}i\rangle$ by

$$\langle\langle C(i,f)\exp(-\Delta S_{\mathrm{F}}/k_B)\rangle\rangle_{\mathrm{F}} = \langle\langle C(i,f)\rangle\rangle_{\mathrm{R}}, \tag{5.1}$$

where $k_{\mathrm{B}}$ is the Boltzmann constant, the forward process average is defined by

$$\left\langle\left\langle C\exp\left(-\Delta S_{\mathrm{F}}/k_B\right)\right\rangle\right\rangle_{\mathrm{F}} \equiv \sum_{i,f} C(i,f)\exp\left(-\Delta S_{\mathrm{F}}/k_B\right)P_{\mathrm{F}}\left(f|i\right) \tag{5.2}$$

with

$$P_{\mathrm{F}}\left(f|i\right) \equiv \left|\langle f|U|i\rangle\right|^2 \rho_{\mathrm{in}}(i), \tag{5.3}$$

where $\tilde{\rho}_{\mathrm{in}}|i\rangle = \rho_{\mathrm{in}}(i)|i\rangle$. The backward process average is defined by

$$\langle\langle C\rangle\rangle_{\mathrm{R}} = \sum_{{}^{\Theta}i,{}^{\Theta}f} C(i,f)P_{\mathrm{R}}\left({}^{\Theta}i|{}^{\Theta}f\right) \tag{5.4}$$

with

$$P_{\mathrm{R}}\left({}^{\Theta}i|{}^{\Theta}f\right) \equiv \left|\left\langle {}^{\Theta}i\right|{}^{\Theta}U\left|{}^{\Theta}f\right\rangle\right|^2 {}^{\Theta}\rho_{\mathrm{fin}}\left({}^{\Theta}f\right). \tag{5.5}$$

For any density matrix $\tilde{\rho}_{\mathrm{fin}}$ for the final eigenstate $|f\rangle$ of the total system, we can show that an operator ${}^{\Theta}\tilde{\rho}_{\mathrm{fin}}$ defined by ${}^{\Theta}\tilde{\rho}_{\mathrm{fin}} \equiv \Theta\tilde{\rho}_{\mathrm{fin}}\Theta^{\dagger}$ is a density matrix for $\left|{}^{\Theta}f\right\rangle$. We then define ${}^{\Theta}\rho_{\mathrm{fin}}\left({}^{\Theta}f\right)$ as an eigenvalue of ${}^{\Theta}\tilde{\rho}_{\mathrm{fin}}$ satisfying $\left|{}^{\Theta}f\right\rangle$ ${}^{\Theta}\tilde{\rho}_{\mathrm{fin}}\left|{}^{\Theta}f\right\rangle = {}^{\Theta}\rho_{\mathrm{fin}}\left({}^{\Theta}f\right)\left|{}^{\Theta}f\right\rangle$ so that ${}^{\Theta}\rho_{\mathrm{fin}}\left({}^{\Theta}f\right)$ is a statistical distribution for $\left|{}^{\Theta}f\right\rangle$. We can also show ${}^{\Theta}\rho_{\mathrm{fin}}\left({}^{\Theta}f\right) = \rho_{\mathrm{fin}}(f)$ since ${}^{\Theta}\tilde{\rho}_{\mathrm{fin}}\left|{}^{\Theta}f\right\rangle = \rho_{\mathrm{fin}}(f)\left|{}^{\Theta}f\right\rangle$, where we have used $\tilde{\rho}_{\mathrm{fin}}|f\rangle = \rho_{\mathrm{fin}}(f)|f\rangle$. Note that we can choose $\tilde{\rho}_{\mathrm{fin}}$ almost freely as long as it can serve as a density matrix for all the eigenstates of $H^{\mathrm{s}}(\tau) + H^{\mathrm{r}}$, $\tilde{N}^{\mathrm{s}}$, and $\tilde{N}^{\mathrm{r}}$.

$\Delta S_{\mathrm{F}}$, which we will call the total entropy production, is defined by

$$\exp\left(-\Delta S_{\mathrm{F}}/k_B\right) \equiv \frac{\rho_{\mathrm{fin}}(f)}{\rho_{\mathrm{in}}(i)} \tag{5.6}$$

so that

$$\Delta S_{\mathrm{F}} = -k_B \ln\left[\frac{\rho_{\mathrm{fin}}(f)}{\rho_{\mathrm{in}}(i)}\right]. \tag{5.7}$$

We can readily show the CKR using

$$P_{\mathrm{R}}\left({}^{\Theta}i\middle|{}^{\Theta}f\right)=\exp\left(-\Delta S_{\mathrm{F}}/k_B\right)P_{\mathrm{F}}\left(f\middle|i\right), \tag{5.8}$$

which follows directly from $\left|\langle f|U|i\rangle\right|^2=\left|\left\langle {}^{\Theta}i\middle|{}^{\Theta}U\middle|{}^{\Theta}f\right\rangle\right|^2$ and ${}^{\Theta}\rho_{\mathrm{fin}}\left({}^{\Theta}f\right)=\rho_{\mathrm{fin}}(f)$.

### B. The total entropy production for a system attached to a heat reservoir

The above definition for the total entropy production $\Delta S_{\mathrm{F}}$ is rather formal. To show that the above definition leads to a more familiar expression for $\Delta S_{\mathrm{F}}$, let us consider a system that exchanges only heat with a heat reservoir at an inverse temperature $\beta$.

Before $t=0$, we detach the system from the reservoir by setting the coupling between the system and the reservoir $H^{\mathrm{int}}$ to be zero and keep the system in equilibrium with a heat reservoir whose inverse temperature is set at $\beta$ while keeping the reservoir in equilibrium with another heat reservoir whose inverse temperature is also set at $\beta$.

Just before $t=0$, through a measurement of the energy in the system, we find the system to be in an initial state $\left|i^{\mathrm{s}}\right\rangle$, which is selected from the eigenstates of the system Hamiltonian $H^{\mathrm{s}}(0)$ by an initial density matrix $\tilde{\rho}^{\mathrm{s}}_{\mathrm{in}}$ corresponding to a canonical ensemble specified by $\beta$.

Through a measurement of the energy in the reservoir, we also find the reservoir to be in an initial state $\left|i^{\mathrm{r}}\right\rangle$, which is selected from the eigenstates of the reservoir Hamiltonian $H^{\mathrm{r}}$ by an initial density matrix $\tilde{\rho}^{\mathrm{r}}_{\mathrm{in}}$ corresponding to a canonical ensemble specified by $\beta$.

The initial state $|i\rangle$ for the total system is then $|i\rangle=\left|i^{\mathrm{s}}\right\rangle\left|i^{\mathrm{r}}\right\rangle$ and the initial density matrix $\tilde{\rho}_{\mathrm{in}}$ for the total system is $\tilde{\rho}_{\mathrm{in}}=\tilde{\rho}^{\mathrm{s}}_{\mathrm{in}}\tilde{\rho}^{\mathrm{r}}_{\mathrm{in}}$ so that the corresponding statistical distribution for the initial state for the total system is

$$\rho_{\mathrm{in}}(i)=\exp\left[\beta\left\{F^{\mathrm{s}}_{\mathrm{in}}-E^{\mathrm{s}}\left(i^{\mathrm{s}}\right)\right\}\right]\exp\left[\beta\left\{F^{\mathrm{r}}-E^{\mathrm{r}}\left(i^{\mathrm{r}}\right)\right\}\right], \tag{5.9}$$

where $E^{\mathrm{s}}\left(i^{\mathrm{s}}\right)$ and $E^{\mathrm{r}}\left(i^{\mathrm{r}}\right)$ are eigenvalues of $H^{\mathrm{s}}(0)$ and $H^{\mathrm{r}}$, respectively, while $F^{\mathrm{s}}_{\mathrm{in}}$ and $F^{\mathrm{r}}$ are the corresponding Helmholtz free energies.

Just after $t = \tau$, we detach the system from the reservoir by setting the coupling between the system and the reservoir $H^{\mathrm{int}}$ to be zero. Through a measurement of the energy in the system, we then find the system to be in a final state $|f^{\mathrm{s}}\rangle$, which is an eigenstate of the system Hamiltonian $H^{\mathrm{s}}(\tau)$. Through a measurement of the energy in the reservoir, we also find the reservoir to be in a final state $|f^{\mathrm{r}}\rangle$, which is an eigenstate of the reservoir Hamiltonian $H^{\mathrm{r}}$.

The final state $|f\rangle$ for the total system is then $|f\rangle = |f^{\mathrm{s}}\rangle|f^{\mathrm{r}}\rangle$. We choose $\rho_{\mathrm{fin}}$ to be the following canonical distribution:

$$\rho_{\mathrm{fin}}(f) = \exp\left[\beta\left\{F^{\mathrm{s}}_{\mathrm{fin}} - E^{\mathrm{s}}\left(f^{\mathrm{s}}\right)\right\}\right]\exp\left[\beta\left\{F^{\mathrm{r}} - E^{\mathrm{r}}\left(f^{\mathrm{r}}\right)\right\}\right], \tag{5.10}$$

where $E^{\mathrm{s}}\left(f^{\mathrm{s}}\right)$ and $E^{\mathrm{r}}\left(f^{\mathrm{r}}\right)$ are eigenvalues of $H^{\mathrm{s}}(\tau)$ and $H^{\mathrm{r}}$, respectively, while $F^{\mathrm{s}}_{\mathrm{fin}}$ is the Helmholtz free energy corresponding to $H^{\mathrm{s}}(\tau)$.

Assuming a *weak* coupling between the system and the reservoir, we define the work done on the system by

$$W \equiv \left\{E^{\mathrm{s}}\left(f^{\mathrm{s}}\right) + E^{\mathrm{r}}\left(f^{\mathrm{r}}\right)\right\} - \left\{E^{\mathrm{s}}\left(i^{\mathrm{s}}\right) + E^{\mathrm{r}}\left(i^{\mathrm{r}}\right)\right\} \tag{5.11}$$

and find the total entropy production to be

$$\exp\left(-\Delta S_{\mathrm{F}} / k_B\right) = \rho_{\mathrm{fin}}(f) / \rho_{\mathrm{in}}(i) = \exp\left(\beta \Delta F^{\mathrm{s}}\right)\exp(-\beta W), \tag{5.12}$$

where $\Delta F^{\mathrm{s}} \equiv F^{\mathrm{s}}_{\mathrm{fin}} - F^{\mathrm{s}}_{\mathrm{in}}$. Using the integral fluctuation theorem, $\left\langle\left\langle \exp\left(-\Delta S_{\mathrm{F}} / k_B\right)\right\rangle\right\rangle_{\mathrm{F}} = 1$, which will be shown in Sec.VIII, this leads to the following Jarzynski equality for the total system:

$$\left\langle\left\langle \exp(-\beta W)\right\rangle\right\rangle_{\mathrm{F}} = \exp\left(-\beta \Delta F^{\mathrm{s}}\right). \tag{5.13}$$

## VI. The Crooks transient fluctuation theorem

The Crooks transient fluctuation theorem,

$$\frac{p_{\mathrm{R}}(\Delta S_{\mathrm{R}} = -\Sigma)}{p_F(\Delta S_F = \Sigma)} = \exp(-\Sigma/k_B), \tag{6.1}$$

which relates the probability $p_F$ for $\Delta S_{\mathrm{F}}$ to take a value of $\Sigma$ after a forward process to the probability $p_{\mathrm{R}}$ for the total entropy production for the backward process,

$$\Delta S_{\mathrm{R}} \equiv -k_B \ln\left[\frac{{}^{\Theta}\rho_{\mathrm{in}}({}^{\Theta}i)}{{}^{\Theta}\rho_{\mathrm{fin}}({}^{\Theta}f)}\right] = -\Delta S_{\mathrm{F}}, \tag{6.2}$$

to take a value of $-\Sigma$, follows from Eq. (5.1) with $C = \delta(\Delta S_{\mathrm{F}} - \Sigma)$ as shown by Crooks for classical systems [7]. If a quantity $C(i,f)$ in Eq. (5.1) is a function of $\Delta S_{\mathrm{F}}$ so that $C = C(\Delta S_{\mathrm{F}})$, then we can rewrite Eq. (5.1) as

$$\sum_{\Sigma} C(\Sigma)\exp(-\Sigma/k_B)p_F(\Delta S_F = \Sigma) = \sum_{\Sigma} C(\Sigma)p_{\mathrm{R}}(\Delta S_{\mathrm{R}} = -\Sigma). \tag{6.3}$$

## VII. The Kawasaki non-linear response relation

The Kawasaki non-linear response relation for an observable $A$ (Eq. (18) in [12 (b)]) used by Andrieux and Gaspard to recover the standard linear response theory can be generalized as

$$\left\langle A_{\mathrm{F}}(\tau)\exp\left(-\Delta\tilde{S}/k_{\mathrm{B}}\right)\right\rangle_{\mathrm{F}} = \left\langle {}^{\Theta}A\right\rangle_{\mathrm{R}}, \tag{7.1}$$

where $\langle A\rangle_{\mathrm{F}} \equiv \mathrm{Tr}[\tilde{\rho}_{\mathrm{in}}A]$, $\langle A\rangle_{\mathrm{R}} \equiv \mathrm{Tr}[{}^{\Theta}\tilde{\rho}_{\mathrm{fin}}A]$, $A_{\mathrm{F}}(\tau) \equiv U^{\dagger}AU$, ${}^{\Theta}A \equiv \Theta A\Theta^{\dagger}$, and

$$\exp\left(-\Delta\tilde{S}/k_B\right) \equiv \left(U^{\dagger}\tilde{\rho}_{\mathrm{fin}}U\right)\left(1/\tilde{\rho}_{\mathrm{in}}\right). \tag{7.2}$$

We can derive this relation using the CKR with

$$C(i,f)=\frac{\langle i|U^{\dagger}A|f\rangle}{\langle i|U^{\dagger}|f\rangle} \tag{7.3}$$

and noting

$$\langle\langle C\exp(-\Delta S_{\mathrm{F}}/k_B)\rangle\rangle_{\mathrm{F}}=\langle A_{\mathrm{F}}(\tau)\exp(-\Delta\tilde{S}/k_{\mathrm{B}})\rangle_{\mathrm{F}} \tag{7.4}$$

and $\langle\langle C\rangle\rangle_{\mathrm{R}}=\langle{}^{\Theta}A\rangle_{\mathrm{R}}$. As this $C(i,f)$ is not a function of $\Delta S_{\mathrm{F}}$, the Kawasaki non-linear response relation cannot be derived from the Crooks transient fluctuation theorem.

## VIII. The integral fluctuation theorem and the Jarzynski equality

The integral fluctuation theorem,

$$\langle\langle\exp(-\Delta S_{\mathrm{F}}/k_B)\rangle\rangle_{\mathrm{F}}=1, \tag{8.1}$$

also follows from Eq. (5.1) with $C=1$ as shown by Crooks for classical systems [7] or from the Crooks transient fluctuation theorem [9 (a), 9 (c) – (e), 10 (a)] as

$$\sum_{\Sigma}\exp(-\Sigma/k_B)p_F(\Delta S_F=\Sigma)=\sum_{\Sigma}p_{\mathrm{R}}(\Delta S_{\mathrm{R}}=-\Sigma)=1 \tag{8.2}$$

or from the Kawasaki non-linear response relation with $A=I$.

All the results obtained so far are independent of particular forms of $\rho_{\mathrm{in}}(i)$ and $\rho_{\mathrm{fin}}(f)$ as long as they can serve as density matrices for the initial and final eigenstates of the total system. For each particular choice of $\rho_{\mathrm{in}}(i)$ and $\rho_{\mathrm{fin}}(f)$, we can then derive *a Jarzynski equality* from the integral fluctuation theorem as shown in [9 (d), 9 (e)] or from the Crooks transient fluctuation theorem as shown in [9 (h), 9 (i), 9 (k), 11 (c)].

## IX. The CKR for steady heat and particle conduction states

### A. The total Hamiltonian

Consider a system driven to a steady state with constant currents of heat and particles by its contact with two reservoirs, A and B, at different inverse temperatures ($\beta^{\mathrm{B}} > \beta^{\mathrm{A}}$) and chemical potentials ($\mu^{\mathrm{A}} > \mu^{\mathrm{B}}$). The total system is subject to a constant magnetic field **B** and the total Hamiltonian is defined by

$$H_{\mathbf{B}} \equiv H^{\mathrm{s}}_{\mathbf{B}} + H^{\mathrm{A}}_{\mathbf{B}} + H^{\mathrm{B}}_{\mathbf{B}} + H^{\mathrm{int}} = H^{(0)}_{\mathbf{B}} + H^{\mathrm{int}}, \qquad (9.1)$$

where $H^{\mathrm{s}}_{\mathbf{B}}$ and $H^{k}_{\mathbf{B}}$ are the Hamiltonians for the system and the *k*-th reservoir while $H^{\mathrm{int}}$ is a *weak* coupling between the system and the reservoirs. Before $t = 0$ and after $t = \tau$, we set $H^{\mathrm{int}} = 0$ so that the system is detached from the reservoirs. We assume that $H_{\mathbf{B}}$ satisfies

$$^{\Theta}H_{\mathbf{B}} = \Theta H_{\mathbf{B}} \Theta^{\dagger} = H_{-\mathbf{B}} \qquad (9.2)$$

so that

$$^{\Theta}U_{\mathbf{B}} = \Theta U_{\mathbf{B}}{}^{\dagger} \Theta^{\dagger} = U_{-\mathbf{B}}. \qquad (9.3)$$

We define the total particle number operator by

$$\tilde{N} \equiv \tilde{N}^{\mathrm{s}} + \tilde{N}^{\mathrm{A}} + \tilde{N}^{\mathrm{B}}, \qquad (9.4)$$

where $\tilde{N}^{\mathrm{s}}$ is the particle number operator for the system and $\tilde{N}^{k}$ is the particle number operator for the $k$-th reservoir We assume that $H_{\mathbf{B}}$ and $\tilde{N}$ commute so that the eigenvalue of $\tilde{N}$ remains constant after each forward process.

### B. The initial eigenstate

This total system has been discussed in Sec.III.B and Sec.III.C as an example. Just before $t = 0$, the system is detached from the reservoirs and through a measurement of both the energy and the number of particles in the system, we find the system to be in an initial state $\left|i_{\mathbf{B}}^{\mathrm{s}}\right\rangle$, which is selected from the eigenstates of the system Hamiltonian $H_{\mathbf{B}}^{\mathrm{s}}$ and $\tilde{N}^{\mathrm{s}}$ by an initial density matrix $\tilde{\rho}_{\mathbf{B}}^{\mathrm{s}}$ corresponding to a grand canonical ensemble specified by $\overline{\beta}$ and $\overline{\mu}$, which we choose to be

$$\overline{\beta} \equiv \frac{\beta^{\mathrm{A}} + \beta^{\mathrm{B}}}{2} \tag{9.5}$$

and

$$\overline{\mu} \equiv \frac{\mu^{\mathrm{A}} + \mu^{\mathrm{B}}}{2}. \tag{9.6}$$

We choose these values for the initial inverse temperature and chemical potential for the system so that we can later simplify the expression for the total entropy production for the total system. Note that after a long time interval, the steady state for the system should be independent of the choice of its initial equilibrium state. We then find the statistical distribution for the initial eigenstate of the system to be

$$\rho_{\mathbf{B}}^{\mathrm{s}}\left(i_{\mathbf{B}}^{\mathrm{s}}\right) = \exp\left[\overline{\beta}\left\{\Omega_{\mathbf{B}}^{\mathrm{s}} - E_{\mathbf{B}}^{\mathrm{s}}\left(i_{\mathbf{B}}^{\mathrm{s}}\right) + \overline{\mu} N^{\mathrm{s}}\left(i_{\mathbf{B}}^{\mathrm{s}}\right)\right\}\right], \tag{9.7}$$

where $\Omega_{\mathbf{B}}^{\mathrm{s}}$ is the thermodynamic potential for the system while $E_{\mathbf{B}}^{\mathrm{s}}\left(i_{\mathbf{B}}^{\mathrm{s}}\right)$ and $N^{\mathrm{s}}\left(i_{\mathbf{B}}^{\mathrm{s}}\right)$ are the initial eigenvalues of $H_{\mathbf{B}}^{\mathrm{s}}$ and $\tilde{N}^{\mathrm{s}}$ for the system.

Just before $t = 0$, through a measurement of both the energy and the number of particles in the reservoirs, we find the $k$-th reservoir to be in an initial state $\left| i_{\mathbf{B}}^{k} \right\rangle$, which is selected from the eigenstates of the reservoir Hamiltonian $H_{\mathbf{B}}^{k}$ and $\tilde{N}^{k}$ by an initial density matrix $\tilde{\rho}_{\mathbf{B}}^{k}$ corresponding to a grand canonical ensemble specified by $\beta^{k}$ and $\mu^{k}$. We then find the statistical distribution for the initial eigenstate of the $k$-th reservoir ($k$ = A, B) to be

$$\rho_{\mathbf{B}}^{k}\left(i_{\mathbf{B}}^{k}\right) = \exp\left[\beta^{k}\left\{\Omega_{\mathbf{B}}^{k} - E_{\mathbf{B}}^{k}\left(i_{\mathbf{B}}^{k}\right) + \mu^{k} N^{k}\left(i_{\mathbf{B}}^{k}\right)\right\}\right], \tag{9.8}$$

where $\Omega_{\mathbf{B}}^{k}$ is the thermodynamic potential for the reservoir while $E_{\mathbf{B}}^{k}\left(i_{\mathbf{B}}^{k}\right)$ and $N^{k}\left(i_{\mathbf{B}}^{k}\right)$ are the initial eigenvalues of $H_{\mathbf{B}}^{k}$ and $N^{k}$ for the reservoir.

The initial eigenstate $\left| i_{\mathbf{B}} \right\rangle$ for the total system is then $\left| i_{\mathbf{B}} \right\rangle = \left| i_{\mathbf{B}}^{\mathrm{s}} \right\rangle \left| i_{\mathbf{B}}^{\mathrm{A}} \right\rangle \left| i_{\mathbf{B}}^{\mathrm{B}} \right\rangle$ and the initial density matrix $\tilde{\rho}_{\mathrm{in}}$ for the total system is $\tilde{\rho}_{\mathrm{in}} = \tilde{\rho}_{\mathbf{B}}^{\mathrm{s}} \tilde{\rho}_{\mathbf{B}}^{\mathrm{A}} \tilde{\rho}_{\mathbf{B}}^{\mathrm{B}} \equiv \tilde{\rho}_{\mathbf{B}}$, which then satisfies

$$^{\Theta}\tilde{\rho}_{\mathbf{B}} = \Theta \tilde{\rho}_{\mathbf{B}} \Theta^{\dagger} = \tilde{\rho}_{-\mathbf{B}}. \tag{9.9}$$

## C. The final eigenstate

Just after $t = \tau$, the system is detached from the reservoirs and through a measurement of both the energy and the number of particles in the system and those in the reservoirs, we find the system to be in an eigenstate $\left| f_{\mathbf{B}}^{\mathrm{s}} \right\rangle$ of the system Hamiltonian $H_{\mathbf{B}}^{\mathrm{s}}$ and $\tilde{N}^{\mathrm{s}}$ while we find the $k$-th reservoir to be in an eigenstate $\left| f_{\mathbf{B}}^{k} \right\rangle$ of the reservoir Hamiltonian $H^{k}$ and $\tilde{N}^{k}$. The final state $\left| f_{\mathbf{B}} \right\rangle$ for the total system is then $\left| f_{\mathbf{B}} \right\rangle = \left| f_{\mathbf{B}}^{\mathrm{s}} \right\rangle \left| f_{\mathbf{B}}^{\mathrm{A}} \right\rangle \left| f_{\mathbf{B}}^{\mathrm{B}} \right\rangle$. We choose the density matrix $\tilde{\rho}_{\mathrm{fin}}$ for the final eigenstate of the total system to be

$$\tilde{\rho}_{\mathrm{fin}} = \tilde{\rho}_{\mathbf{B}} = \tilde{\rho}_{\mathbf{B}}^{\mathrm{s}} \tilde{\rho}_{\mathbf{B}}^{\mathrm{A}} \tilde{\rho}_{\mathbf{B}}^{\mathrm{B}} \tag{9.10}$$

so that $\rho_{\mathbf{B}}\left(f_{\mathbf{B}}\right) = \rho_{\mathbf{B}}^{\mathrm{s}}\left(f_{\mathbf{B}}^{\mathrm{s}}\right) \rho_{\mathbf{B}}^{\mathrm{A}}\left(f_{\mathbf{B}}^{\mathrm{A}}\right) \rho_{\mathbf{B}}^{\mathrm{B}}\left(f_{\mathbf{B}}^{\mathrm{B}}\right)$.

**D. The Crooks-Kawasaki relation**

The CKR then becomes

$$\langle\langle C(i_{\mathbf{B}},f_{\mathbf{B}})\exp(-\Delta S_{\mathrm{F}}/k_B)\rangle\rangle_{\mathrm{F},\mathbf{B}}=\langle\langle C(f_{\mathbf{B}},i_{\mathbf{B}})\rangle\rangle_{\mathrm{F},-\mathbf{B}}, \tag{9.11}$$

where

$$\langle\langle C(f_{\mathbf{B}},i_{\mathbf{B}})\rangle\rangle_{\mathrm{F},-\mathbf{B}}=\sum_{i_{-\mathbf{B}},f_{-\mathbf{B}}}C(f_{\mathbf{B}},i_{\mathbf{B}})P_{\mathrm{F},-\mathbf{B}}(f_{-\mathbf{B}}|i_{-\mathbf{B}}) \tag{9.12}$$

as $P_{\mathrm{R},-\mathbf{B}}=P_{\mathrm{F},-\mathbf{B}}$.

By applying the first-order time-dependent perturbation theory, where we assume $H^{\mathrm{int}}$ to be weak, we find that $|\langle f_{\mathbf{B}}|U_{\mathbf{B}}|i_{\mathbf{B}}\rangle|^2$ is appreciable only when

$$\left|\left\{E_{\mathbf{B}}^{\mathrm{s}}(f_{\mathbf{B}}^{\mathrm{s}})+E_{\mathbf{B}}^{\mathrm{A}}(f_{\mathbf{B}}^{\mathrm{A}})+E_{\mathbf{B}}^{\mathrm{B}}(f_{\mathbf{B}}^{\mathrm{B}})\right\}-\left\{E_{\mathbf{B}}^{\mathrm{s}}(i_{\mathbf{B}}^{\mathrm{s}})+E_{\mathbf{B}}^{\mathrm{A}}(i_{\mathbf{B}}^{\mathrm{A}})+E_{\mathbf{B}}^{\mathrm{B}}(i_{\mathbf{B}}^{\mathrm{B}})\right\}\right|<\frac{2\pi}{\tau}. \tag{9.13}$$

For sufficiently long $\tau$, we can then assume

$$W\equiv\Delta E_{\mathbf{B}}^{\mathrm{s}}+\Delta E_{\mathbf{B}}^{\mathrm{A}}+\Delta E_{\mathbf{B}}^{\mathrm{B}}=0, \tag{9.14}$$

where we define the work $W$ done on the system using $\Delta E_{\mathbf{B}}^{\mathrm{s}}\equiv E_{\mathbf{B}}^{\mathrm{s}}(f_{\mathbf{B}}^{\mathrm{s}})-E_{\mathbf{B}}^{\mathrm{s}}(i_{\mathbf{B}}^{\mathrm{s}})$ and $\Delta E_{\mathbf{B}}^{k}\equiv E_{\mathbf{B}}^{k}(f_{\mathbf{B}}^{k})-E_{\mathbf{B}}^{k}(i_{\mathbf{B}}^{k})$. Using $W=0$ and $N(f_{\mathbf{B}})=N(i_{\mathbf{B}})$, we find

$$\exp(-\Delta S_{\mathrm{F}}/k_B)=\exp\left[-\tau\left(\Delta\beta J_{\mathbf{B}}^{Q}+\bar{\beta}\Delta\mu J_{\mathbf{B}}^{N}\right)\right], \tag{9.15}$$

where $\Delta\beta\equiv\beta^{B}-\beta^{A}$ and $\Delta\mu\equiv\mu^{A}-\mu^{B}$. The average heat and particle currents are defined by

$$J_{\mathbf{B}}^{Q}\equiv\frac{Q_{\mathbf{B}}^{\mathrm{A}}-Q_{\mathbf{B}}^{\mathrm{B}}}{2\tau} \tag{9.16}$$

and

$$J_{\mathbf{B}}^{N} \equiv \frac{\Delta N_{\mathbf{B}}^{\mathrm{B}} - \Delta N_{\mathbf{B}}^{\mathrm{A}}}{2\tau}, \tag{9.17}$$

where the heat transferred from the *k*-th reservoir into the system is defined by

$$Q_{\mathbf{B}}^{k} \equiv -\left(\Delta E_{\mathbf{B}}^{k} - \mu^{k} \Delta N_{\mathbf{B}}^{k}\right) \tag{9.18}$$

and the number of particles transferred from the *k*-th reservoir into the system is defined by

$$\Delta N_{\mathbf{B}}^{k} \equiv N^{k}\left(f_{\mathbf{B}}^{k}\right) - N^{k}\left(i_{\mathbf{B}}^{k}\right), \tag{9.19}$$

so that $\Delta E_{\mathbf{B}}^{\mathrm{s}}$ satisfies

$$\Delta E_{\mathbf{B}}^{\mathrm{s}} = -\Delta E_{\mathbf{B}}^{\mathrm{A}} - \Delta E_{\mathbf{B}}^{\mathrm{B}} = \left(Q_{\mathbf{B}}^{\mathrm{A}} + Q_{\mathbf{B}}^{\mathrm{B}}\right) + W_{\mathbf{B}}^{\mathrm{c}}, \tag{9.20}$$

where $W_{\mathbf{B}}^{\mathrm{c}} \equiv \mu^{A}\left(-\Delta N_{\mathbf{B}}^{\mathrm{A}}\right) + \mu^{B}\left(-\Delta N_{\mathbf{B}}^{\mathrm{B}}\right)$ is the "chemical work" done on the system by transfer of particles from the reservoirs.

## X. The fluctuation theorem for the currents

We define the cumulant generating function for the currents by

$$G\left(\{\lambda_{Q}, \lambda_{N}\}; \{\Delta\beta, \bar{\beta}\Delta\mu\}; \mathbf{B}\right) \equiv -\lim_{\tau\to\infty} \frac{1}{\tau} \ln\left\langle\left\langle \exp\left[-\tau\left(\lambda_{Q} J_{\mathbf{B}}^{Q} + \lambda_{N} J_{\mathbf{B}}^{N}\right)\right]\right\rangle\right\rangle_{\mathrm{F},\mathbf{B}}. \tag{10.1}$$

Substituting $C = \exp\left[\tau\left(\lambda_{Q} J_{\mathbf{B}}^{Q} + \lambda_{N} J_{\mathbf{B}}^{N}\right)\right]$ into the CKR, we can prove the fluctuation theorem for the currents:

$$G\left(\left\{\Delta\beta-\lambda_Q,\bar{\beta}\Delta\mu-\lambda_N\right\},\left\{\Delta\beta,\bar{\beta}\Delta\mu\right\},\mathbf{B}\right)=G\left(\left\{\lambda_Q,\lambda_N\right\},\left\{\Delta\beta,\bar{\beta}\Delta\mu\right\},-\mathbf{B}\right) \quad (10.2)$$

because

$$\left\langle\left\langle \exp\left[\tau\left(\lambda_Q J_{\mathbf{B}}^Q+\lambda_N J_{\mathbf{B}}^N\right)\right]\exp\left[-\tau\left(\Delta\beta J_{\mathbf{B}}^Q+\bar{\beta}\Delta\mu J_{\mathbf{B}}^N\right)\right]\right\rangle\right\rangle_{\mathrm{F},\mathbf{B}}=\left\langle\left\langle \exp\left[-\tau\left(\lambda_Q J_{-\mathbf{B}}^Q+\tau\lambda_N J_{-\mathbf{B}}^Q\right)\right]\right\rangle\right\rangle_{\mathrm{F},-\mathbf{B}}. \quad (10.3)$$

As both $C$ and $\Delta S_{\mathrm{F}}=k_B\tau\left(\Delta\beta J_{\mathbf{B}}^Q+\bar{\beta}\Delta\mu J_{\mathbf{B}}^N\right)$ are functions of $J_{\mathbf{B}}^Q$ and $J_{\mathbf{B}}^N$, we can also obtain this relation from the following variant of the Crooks transient fluctuation theorem:

$$\frac{p_{\mathrm{F},-\mathbf{B}}\left(J_{\mathbf{B}}^Q=-\hat{J}_{\mathbf{B}}^Q,J_{\mathbf{B}}^N=-\hat{J}_{\mathbf{B}}^N\right)}{p_{\mathrm{F},\mathbf{B}}\left(J_{\mathbf{B}}^Q=\hat{J}_{\mathbf{B}}^Q,J_{\mathbf{B}}^N=\hat{J}_{\mathbf{B}}^N\right)}=\exp\left[-\Delta S_{\mathrm{F}}\left(\hat{J}_{\mathbf{B}}^Q,\hat{J}_{\mathbf{B}}^N\right)/k_B\right]. \quad (10.4)$$

Using the CKR, we can also prove the fluctuation theorem for the currents for the generating function for average energy and particle currents [12 (c)] and for the generating function used in full counting statistics [12 (a)].

## XI. The Green-Kubo relation

The forward process average of the $i$-th current $(i=Q,N)$ is obtained by

$$\left\langle\left\langle J_{\mathbf{B}}^i\right\rangle\right\rangle_{\mathrm{F},\mathbf{B}}=\left.\frac{\partial G}{\partial\lambda_i}\right|_{\lambda_Q=\lambda_N=0}. \quad (11.1)$$

For the affinities defined by $a_Q\equiv\Delta\beta$ and $a_N\equiv\bar{\beta}\Delta\mu$, we then find the linear response coefficients by

$$L_{ij}(\mathbf{B}) = \left.\frac{\partial J_i(\mathbf{B})}{\partial a_j}\right|_{a_Q=a_N=0} = \left.\frac{\partial^2 G}{\partial a_j \partial \lambda_i}\right|_{\lambda_Q=\lambda_N=a_Q=a_N=0} \qquad (i,j=Q,N), \quad (11.2)$$

where

$$J_i(\mathbf{B}) \equiv \lim_{\tau\to\infty} \left\langle\left\langle J_{\mathbf{B}}^i \right\rangle\right\rangle_{F,\mathbf{B}} \cong L_{ii}(\mathbf{B}) a_i + L_{ij}(\mathbf{B}) a_j . \quad (11.3)$$

Using the fluctuation theorem for the currents and its corollary,

$$G\left(\{\Delta\beta, \bar{\beta}\Delta\mu\}, \{\Delta\beta, \bar{\beta}\Delta\mu\}, \mathbf{B}\right) = 0, \quad (11.4)$$

we can then derive [12 (c)] the Green-Kubo relation,

$$L_{ii}(\mathbf{B}) = L_{ii}(-\mathbf{B}) = \lim_{\tau\to\infty} \frac{\tau}{2} \left\langle\left\langle \left(J_{\mathbf{B}}^i\right)^2 \right\rangle\right\rangle_{\mathrm{eq},\mathbf{B}}, \quad (11.5)$$

where "eq" indicates that the average is taken with $\Delta\beta = \bar{\beta}\Delta\mu = 0$.

## XII. The Green-Kubo formula for thermal conductivity

Consider a system that is not subject to a magnetic field and exchanges heat with two reservoirs at two temperatures satisfying $T^{\mathrm{A}} > T^{\mathrm{B}}$. The linear response relation for the average heat current $J_Q$ is

$$J_Q = \kappa A \frac{T^{\mathrm{A}} - T^{\mathrm{B}}}{L}, \quad (12.1)$$

where $\kappa$, $A$, and $L$ are the thermal conductivity, the cross-sectional area, and the length of the system. Using the equation for $L_{QQ}$ given above, we find

$$\kappa = \frac{V}{2k_B\overline{T}^2}\lim_{\tau\to\infty}\tau\left\langle\left(j^Q\right)^2\right\rangle_{\rm eq}, \tag{12.2}$$

where $V = AL$,

$$j^Q \equiv \frac{J^Q}{A} = \frac{Q^{\rm A} - Q^{\rm B}}{2A\tau}, \tag{12.3}$$

and

$$\Delta\beta \cong \frac{T^A - T^B}{k_B\overline{T}^2}$$

with $k_{\rm B}\overline{\beta} = 1/\overline{T}$.

Recalling $Q^k = -\Delta E^k = -\left\{E^k(f) - E^k(i)\right\}$, we can then show

$$\left\langle\left(j^Q\right)^2\right\rangle_{\rm eq} = \frac{1}{(2A\tau)^2}{\rm Tr}\left[\rho_{\rm eq}\left[\left\{H_{\rm F}^{\rm A}(\tau) - H_{\rm F}^{\rm A}(0)\right\} - \left\{H_{\rm F}^{\rm B}(\tau) - H_{\rm F}^{\rm B}(0)\right\}\right]^2\right], \tag{12.4}$$

where

$$\tilde{\rho}_{\rm eq} \equiv \exp\left[\overline{\beta}\left(F^{\rm s} - H^{\rm s}\right)\right]\exp\left[\overline{\beta}\left(F^{\rm A} - H^{\rm A}\right)\right]\exp\left[\overline{\beta}\left(F^{\rm B} - H^{\rm B}\right)\right], \tag{12.5}$$

where $F^{\rm s}$ and $F^k$ are the Helmholtz free energies of the system and the $k$-th reservoir in their equilibrium states, and we have also used $\left[H_{\rm F}^{\rm A}(0), H_{\rm F}^{\rm B}(0)\right] = \left[H_{\rm F}^{\rm A}(\tau), H_{\rm F}^{\rm B}(\tau)\right] = 0$ and $\left[\tilde{\rho}_{\rm eq}, H_{\rm F}^k(0)\right] = 0$. We define the heat current density operator in the Heisenberg picture by

$$\tilde{j}_{\rm F}^Q(t) \equiv \left(\frac{1}{2A}\right)\left(\frac{dH_{\rm F}^{\rm A}}{dt} - \frac{dH_{\rm F}^{\rm B}}{dt}\right), \tag{12.6}$$

where

$$i\hbar\frac{dH_{\rm F}^k}{dt} = \left[H_{\rm F}^k(t), H_{\rm F}(t)\right], \tag{12.7}$$

so that

$$\int_0^\tau dt\tilde{j}_{\rm F}^Q(t) = \frac{1}{2A}\left[\left\{H_{\rm F}^{\rm A}(\tau) - H_{\rm F}^{\rm A}(0)\right\} - \left\{H_{\rm F}^{\rm B}(\tau) - H_{\rm F}^{\rm B}(0)\right\}\right]. \tag{12.8}$$

We then obtain

$$\kappa = \frac{V}{2k_B\overline{T}^2}\lim_{\tau\to\infty}\frac{1}{\tau}\mathrm{Tr}\left[\rho_{\mathrm{eq}}\left\{\int_0^\tau dt\tilde{j}_{\mathrm{F}}^Q(t)\right\}^2\right]. \tag{12.9}$$

Assuming $U(t_2)\tilde{\rho}_{\mathrm{eq}}U(t_2)^\dagger = \tilde{\rho}_{\mathrm{eq}}$, where $U(t_2)$ is the time evolution operator at $t_2$, we obtain the Green-Kubo formula for $\kappa$,

$$\kappa = \frac{V}{k_B\overline{T}^2}\int_0^\infty dt\tilde{C}_Q(t), \tag{12.10}$$

where

$$\tilde{C}_Q \equiv \mathrm{Tr}\left[\tilde{\rho}_{\mathrm{eq}}\frac{1}{2}\left\{\tilde{j}_{\mathrm{F}}^Q(t)\tilde{j}_{\mathrm{F}}^Q(0)+\tilde{j}_{\mathrm{F}}^Q(0)\tilde{j}_{\mathrm{F}}^Q(t)\right\}\right] \tag{12.11}$$

is assumed to satisfy

$$\lim_{\tau\to\infty}\frac{1}{\tau}\int_0^\tau dt\tilde{C}_Q(t)t = 0. \tag{12.12}$$

## XIII. Extensions of the Clausius and the Gibbs relations to steady heat conduction states

### A. Generalized entropy and an extension of the second law

For a quantum system driven out of equilibrium, Tasaki [9 (b)] derived

$$\langle\langle \Delta S_{\mathrm{F}}\rangle\rangle_{\mathrm{F}} \geq 0 \tag{13.1}$$

from $\langle\langle \exp(-\Delta S_{\mathrm{F}}/k_B)\rangle\rangle_{\mathrm{F}} = 1$ using Jensen's inequality, $\langle\langle \exp(C)\rangle\rangle_{\mathrm{F}} \geq \exp(\langle\langle C\rangle\rangle_{\mathrm{F}})$, and $e^{-x} \geq 1-x$. Defining a generalized entropy for the final non-equilibrium state by

$$S_{\text{fin}} \equiv -k_{\text{B}} \langle\langle \ln \rho_{\text{fin}}(f) \rangle\rangle_{\text{F}} = -k_{\text{B}} \text{Tr}[\tilde{\rho}(\tau) \ln \tilde{\rho}_{\text{fin}}], \tag{13.2}$$

where $\tilde{\rho}(\tau) = U\tilde{\rho}_{\text{in}} U^{\dagger}$, and noting that the entropy for the initial equilibrium state satisfies

$$S_{\text{eq}} \equiv -k_{\text{B}} \langle\langle \ln \rho_{\text{in}}(i) \rangle\rangle_{\text{F}} = -k_{\text{B}} \text{Tr}[\tilde{\rho}_{\text{in}} \ln \tilde{\rho}_{\text{in}}] = -k_{\text{B}} \text{Tr}[\tilde{\rho}(\tau) \ln \tilde{\rho}(\tau)], \tag{13.3}$$

he found $\langle\langle \Delta S_{\text{F}} \rangle\rangle_{\text{F}} = S_{\text{fin}} - S_{\text{eq}} \geq 0$, which is consistent with $\langle\langle \Delta S_{\text{F}} \rangle\rangle_{\text{F}} / k_{\text{B}}$ being a non-negative relative entropy [9 (b), 14],

$$\langle\langle \Delta S_{\text{F}} \rangle\rangle_{\text{F}} / k_{\text{B}} = \text{Tr}[\tilde{\rho}(\tau) \{\ln \tilde{\rho}(\tau) - \ln \tilde{\rho}_{\text{fin}}\}] \geq 0. \tag{13.4}$$

$S_{\text{fin}} \geq S_{\text{eq}}$ can be considered as an extension of the second law to forward transient processes as Tasaki showed that it is reduced to the second law for an isolated system going through an adiabatic process.

### B. Internal energy of the system and the excess heat from the reservoirs

Consider a system that exchanges heat with two reservoirs whose inverse temperatures are fixed by their average $\overline{\beta}$ and their difference $\Delta\beta$. Suppose the system starts in an equilibrium state at $\overline{\beta} = 1/(k_{\text{B}} \overline{T})$ and, after a sufficiently long time $\tau$, settles into a steady state with a constant average heat current. Both its volume $V$ and its number of particles are kept constant. Assuming a *weak* coupling between the system and the reservoirs, we find

$$0 = \Delta E^{\text{s}} + \Delta E^{\text{A}} + \Delta E^{\text{B}} = \Delta E^{\text{s}} - \left(Q^{\text{A}} + Q^{\text{B}}\right) \tag{13.5}$$

as justified earlier so that

$$U_{\text{st}}^{\text{s}} = U_{\text{eq}}^{\text{s}} + Q, \tag{13.6}$$

where $Q \equiv \langle\langle Q^{\mathrm{A}} + Q^{\mathrm{B}} \rangle\rangle_{\mathrm{F}}$ is the total amount of heat transferred into the system while $U_{\mathrm{st}}^{\mathrm{s}} \equiv \langle\langle E^{\mathrm{s}}(f) \rangle\rangle_{\mathrm{F}}$ and $U_{\mathrm{eq}}^{\mathrm{s}} \equiv \langle\langle E^{\mathrm{s}}(i) \rangle\rangle_{\mathrm{F}}$ are the system's internal energies for its final steady state and for its initial equilibrium state. As $\tau$ increases, the average heat current $\langle\langle J_k^Q \rangle\rangle_{\mathrm{F}}$ from the $k$-th reservoir becomes independent of $\tau$ so that $\lim_{\tau\to\infty} \langle\langle J_k^Q \rangle\rangle_{\mathrm{F}} = J_{\mathrm{st}}^k$. Since $U_{\mathrm{st}}^{\mathrm{s}}$ also becomes independent of $\tau$, so does $Q$. We then find

$$J_{\mathrm{st}}^{\mathrm{A}} + J_{\mathrm{st}}^{\mathrm{B}} = \lim_{\tau\to\infty} \frac{Q}{\tau} = 0 . \tag{13.7}$$

By defining the "excess heat" [15] from the $k$-th reservoir by

$$\langle\langle Q_{\mathrm{ex}}^k \rangle\rangle_{\mathrm{F}} \equiv \langle\langle Q^k \rangle\rangle_{\mathrm{F}} - \tau J_{\mathrm{st}}^k , \tag{13.8}$$

we then obtain

$$Q_{\mathrm{ex}} \equiv \langle\langle Q_{\mathrm{ex}}^{\mathrm{A}} + Q_{\mathrm{ex}}^{\mathrm{B}} \rangle\rangle_{\mathrm{F}} = Q \tag{13.9}$$

so that

$$U_{\mathrm{st}}^{\mathrm{s}} = U_{\mathrm{eq}}^{\mathrm{s}} + Q_{\mathrm{ex}} . \tag{13.10}$$

As $U_{\mathrm{st}}^{\mathrm{s}}$ is a function of $\overline{\beta}$, $\Delta\beta$, and $V$ and therefore it is a state variable for the final steady state, the total excess heat $Q_{\mathrm{ex}}$ is also a state variable.

### C. Generalized entropies of the system and the reservoirs and the excess heat

We can divide $\langle\langle \Delta S_{\mathrm{F}} \rangle\rangle_{\mathrm{F}}$ into two parts, one for the system and the other for the reservoirs: $\langle\langle \Delta S_{\mathrm{F}} \rangle\rangle_{\mathrm{F}} = \Delta S^{\mathrm{s}} + \Delta S^{\mathrm{r}} \geq 0$, where

$$\Delta S^{\mathrm{s}} \equiv -k_{\mathrm{B}} \left\langle\left\langle \ln\left[ \frac{\rho_{\mathrm{fin}}^{\mathrm{s}}(f)}{\rho_{\mathrm{in}}^{\mathrm{s}}(i)} \right] \right\rangle\right\rangle_{\mathrm{F}} \tag{13.11}$$

and

$$\Delta S^{\mathrm{r}} \equiv -k_{\mathrm{B}}\left\langle\left\langle \ln\left[\frac{\rho_{\mathrm{fin}}^{\mathrm{A}}(f)\rho_{\mathrm{fin}}^{\mathrm{B}}(f)}{\rho_{\mathrm{in}}^{\mathrm{A}}(i)\rho_{\mathrm{in}}^{\mathrm{B}}(i)}\right]\right\rangle\right\rangle_{\mathrm{F}}. \tag{13.12}$$

We assume

$$\tilde{\rho}_{\mathrm{in}}^{\mathrm{s}} = \tilde{\rho}_{\mathrm{fin}}^{\mathrm{s}} = \tilde{\rho}^{\mathrm{s}} = \exp\left[\overline{\beta}\left(F^{\mathrm{s}} - H^{\mathrm{s}}\right)\right] \tag{13.13}$$

for the system and

$$\tilde{\rho}_{\mathrm{in}}^{k} = \tilde{\rho}_{\mathrm{fin}}^{k} = \tilde{\rho}^{k} = \exp\left[\beta^{k}\left(F^{k} - H^{k}\right)\right] \tag{13.14}$$

for the $k$-th reservoir. $F^{\mathrm{s}}$ and $F^{k}$ are the Helmholtz free energies of the system and the $k$-th reservoir in their respective initial equilibrium states and satisfy $F^{\mathrm{s}} = U_{\mathrm{eq}}^{\mathrm{s}} - \overline{T}S_{\mathrm{eq}}^{\mathrm{s}}$, where $S_{\mathrm{eq}}^{\mathrm{s}}$ is the entropy of the system in its initial equilibrium state, and $F^{k} = U_{\mathrm{eq}}^{k} - T^{k}S_{\mathrm{eq}}^{k}$, where $T^{k}$, $U_{\mathrm{eq}}^{k}$, and $S_{\mathrm{eq}}^{k}$ are the temperature, the internal energy, and the entropy of the $k$-th reservoir in its initial equilibrium state.

Defining the generalized entropy for the system in its final steady state by $S_{\mathrm{st}}^{\mathrm{s}} \equiv -k_{\mathrm{B}}\left\langle\left\langle \ln\rho^{\mathrm{s}}(f)\right\rangle\right\rangle_{\mathrm{F}}$, we find

$$S_{\mathrm{st}}^{\mathrm{s}} - S_{\mathrm{eq}}^{\mathrm{s}} = \Delta S^{\mathrm{s}} = -k_{\mathrm{B}}\left\langle\left\langle \ln\left[\frac{\exp\left[\overline{\beta}\left(F^{\mathrm{s}} - E^{\mathrm{s}}(f)\right)\right]}{\exp\left[\overline{\beta}\left(F^{\mathrm{s}} - E^{\mathrm{s}}(i)\right)\right]}\right]\right\rangle\right\rangle_{\mathrm{F}} = \frac{U_{\mathrm{st}}^{\mathrm{s}} - U_{\mathrm{eq}}^{\mathrm{s}}}{\overline{T}} = \frac{Q_{\mathrm{ex}}}{\overline{T}} \tag{13.15}$$

so that

$$S_{\mathrm{st}}^{\mathrm{s}} = S_{\mathrm{eq}}^{\mathrm{s}} + \frac{Q_{\mathrm{ex}}}{\overline{T}}, \tag{13.16}$$

where $S_{\mathrm{eq}}^{\mathrm{s}} \equiv -k_{\mathrm{B}}\left\langle\left\langle \ln\rho^{\mathrm{s}}(i)\right\rangle\right\rangle_{\mathrm{F}}$. $S_{\mathrm{st}}^{\mathrm{s}}$ is thus a state variable for the steady state.

For the reservoirs, we also find

$$\Delta S^{\mathrm{r}} = -k_{\mathrm{B}} \sum_k \beta^k \langle\langle Q^k \rangle\rangle_{\mathrm{F}} = -\frac{Q_{\mathrm{ex}}}{\overline{T}} + k_{\mathrm{B}} \tau \Delta\beta \langle\langle J^Q \rangle\rangle_{\mathrm{F}} \tag{13.17}$$

so that

$$\langle\langle \Delta S_{\mathrm{F}} \rangle\rangle_{\mathrm{F}} = k_{\mathrm{B}} \tau \Delta\beta \langle\langle J^Q \rangle\rangle_{\mathrm{F}}. \tag{13.18}$$

We can then obtain the average steady heat current from the rate of the average total entropy production as

$$J_{\mathrm{st}} \equiv \lim_{\tau\to\infty} \langle\langle J^Q \rangle\rangle_{\mathrm{F}} = \frac{1}{k_{\mathrm{B}}\Delta\beta} \lim_{\tau\to\infty} \frac{\langle\langle \Delta S_{\mathrm{F}} \rangle\rangle_{\mathrm{F}}}{\tau}. \tag{13.19}$$

**D. Extensions of the Clausius and the Gibbs relations**

$\Delta S^{\mathrm{s}} = Q_{\mathrm{ex}} / \overline{T}$ is an extension of the Clausius relation, $\Delta S^{\mathrm{s}}_{\mathrm{eq}} = Q^{\mathrm{qs}}_T / T$, for a system that accepts heat $Q^{\mathrm{qs}}_T$ through a quasi-static isothermal process at temperature $T$. With fixed $\overline{T}$ and $V$, the transient process from the equilibrium state with $\Delta\beta = 0$ to the final steady state with $\Delta\beta \neq 0$ is then analogous to a quasi-static isothermal process. More generally, using $S^{\mathrm{s}}_{\mathrm{st}} = S^{\mathrm{s}}_{\mathrm{eq}} + Q_{\mathrm{ex}} / \overline{T}$, we find an extension of the Clausius relation to an infinitesimal process from a steady state at $(\overline{T}, \Delta\beta, V, n)$ to another at $(\overline{T} + d\overline{T}, \Delta\beta + d(\Delta\beta), V + dV, n)$ to be

$$dS^{\mathrm{s}}_{\mathrm{st}} = \frac{\delta Q^{\mathrm{qs}}}{\overline{T}} + d\left(\frac{Q_{\mathrm{ex}}}{\overline{T}}\right), \tag{13.20}$$

where $\delta Q^{\mathrm{qs}}$ is the heat transferred into the system during a quasi-static process from an equilibrium state at $(\overline{T}, V, n)$ to another at $(\overline{T} + d\overline{T}, V + dV, n)$ and satisfies $dS^{\mathrm{s}}_{\mathrm{eq}} = \delta Q^{\mathrm{qs}} / \overline{T}$. Using $dQ_{\mathrm{ex}} = dU^{\mathrm{s}}_{\mathrm{st}} - dU^{\mathrm{s}}_{\mathrm{eq}}$ together with $dU^{\mathrm{s}}_{\mathrm{eq}} = \overline{T} dS^{\mathrm{s}}_{\mathrm{eq}} - P^{\mathrm{s}}_{\mathrm{eq}} dV$, where $P^{\mathrm{s}}_{\mathrm{eq}}$ is the system's equilibrium pressure, we then obtain an extension of the Gibbs relation,

$$dS_{\rm st}^{\rm s} = \frac{1}{\overline{T}} dU_{\rm st}^{\rm s} + \frac{P_{\rm eq}^{\rm s}}{\overline{T}} dV - \frac{Q_{\rm ex}}{\overline{T}^2} d\overline{T} \,. \tag{13.21}$$

## XIV. Steady states driven by work: the shear viscosity of a fluid

Finally, we consider a fluid driven to a steady state with a shear flow. The fluid is placed between two plates both perpendicular to the $y$-axis and each with a surface area $A$ and the depth of the fluid along the $y$-axis is $h$. The fluid is also attached to a heat reservoir at an inverse temperature $\beta$. While the plate under the fluid is at rest, the plate above is moving at a constant velocity $v$ in the $x$-direction to induce the shear flow in the fluid during $[0,\tau]$. On the macroscopic level, the average work done on the fluid by the moving plate is $\overline{W} = A\overline{P}_{yx} v\tau$, where $\overline{P}_{yx}$ is the average shear stress exerted on the fluid by the plate. For small $|v|$, we define the shear viscosity $\eta$ of the fluid by a linear response relation,

$$\overline{P}_{yx} \cong \eta v / h \,. \tag{14.1}$$

The total Hamiltonian is

$$H(t,v) = H^{(0)} + H^{\rm int}\left[\{\mathbf{r}_a + (vt)\hat{x}\}_a\right], \tag{14.2}$$

where $H^{\rm int}$ is the potential energy due to the interactions between the particles in the fluid and those in the moving plate while $H^{(0)}$ is the rest of the total Hamiltonian. $\mathbf{r}_a + (vt)\hat{x}$ is the position vector for the $a$-th atom in the moving plate. For the fluid and the plates, we impose periodic boundary conditions at the boundaries perpendicular to the $x$ axis. We assume that the moving plate returns to its initial position at $t = \tau$ so that $H(0,v) = H(\tau,v)$. We also assume that

$$^{\Theta}H(t,v) \equiv \Theta H(t,v)\Theta^{\dagger} = H(t,-v) \,. \tag{14.3}$$

If we assume

$$\rho_{\rm in}(i) \equiv \exp[\beta\{F - E(i)\}] \tag{14.4}$$

and

$$\rho_{\rm fin}(f) = \exp[\beta\{F - E(f)\}], \tag{14.5}$$

where $E(i)$ and $E(f)$ are the eigenvalues of $H(0,v)$ and $H(\tau,v)$, and define the work done on the system by

$$W \equiv E(f) - E(i) \tag{14.6}$$

and the process-dependent shear stress by

$$P_{yx} \equiv \frac{W}{Av\tau}, \tag{14.7}$$

the CKR becomes

$$\langle\langle C(i,f)\exp(-\Delta S_{\rm F}/k_B)\rangle\rangle_{{\rm F},v} = \langle\langle C(f,\ i)\rangle\rangle_{{\rm F},-v} \tag{14.8}$$

with

$$\exp(-\Delta S_{\rm F}/k_B) = \exp(-\beta W) = \exp(-\tau A\beta v P_{yx}). \tag{14.9}$$

We define the cumulant generating function for the shear stress by

$$G_P(\lambda_P, v) \equiv -\lim_{\tau\to\infty}\frac{1}{\tau}\ln\langle\langle\exp(-\tau\lambda_P P_{yx})\rangle\rangle_{{\rm F},v}, \tag{14.10}$$

with which we obtain the average shear stress by

$$\langle\langle P_{yx} \rangle\rangle_{\mathrm{F}} = \left.\frac{\partial G_P}{\partial \lambda_P}\right|_{\lambda_P = 0} \tag{14.11}$$

and the shear viscosity by

$$\eta = h \left.\frac{\partial \langle\langle P_{yx} \rangle\rangle_{\mathrm{F}}}{\partial v}\right|_{v=0} = h \left.\frac{\partial^2 G_P}{\partial v \partial \lambda_P}\right|_{\lambda_P = v = 0}. \tag{14.15}$$

The fluctuation theorem for the shear stress,

$$G_P(A\beta v - \lambda_P, v) = G_P(-\lambda_P, -v), \tag{14.15}$$

follows from the CKR with $C = \exp(\tau \lambda_P P_{yx})$ because

$$\langle\langle \exp(\tau \lambda_P P_{yx}) \exp(-\tau A \beta v P_{yx}) \rangle\rangle_{\mathrm{F},v} = \langle\langle \exp(\tau \lambda_P P_{yx}) \rangle\rangle_{\mathrm{F},-v}. \tag{14.16}$$

Using $G_P(A\beta v, v) = 0$, which follows from $G_P(A\beta v - \lambda_P, v) = G_P(-\lambda_P, -v)$, we can show

$$\eta = \frac{V}{2k_B T} \lim_{\tau \to \infty} \tau \langle\langle P_{yx}{}^2 \rangle\rangle_{\mathrm{eq}}, \tag{14.17}$$

where $V = Ah$, $k_B T = 1/\beta$, and “eq” indicates that the average is taken with $v = 0$. Using

$$H_{\mathrm{F}}(\tau, v) - H_{\mathrm{F}}(0, v) = v \int_0^\tau dt \sum_a \left( \frac{\partial H^{\mathrm{int}}}{\partial x_a} \right)_{\mathrm{F}} \equiv Av \int_0^\tau dt \tilde{P}_{\mathrm{F}}(t), \tag{14.18}$$

where $H_{\mathrm{F}}(\tau, v) = U^\dagger H(\tau, v) U$ and the shear stress operator $\tilde{P}$ is defined, we obtain

$$\left\langle\left\langle P_{yx}{}^{2}\right\rangle\right\rangle_{\mathrm{eq}} = \mathrm{Tr}\left[\tilde{\rho}_{\mathrm{in}}\left\{\frac{H_{\mathrm{F}}(\tau,v)-H_{\mathrm{F}}(0,v)}{Av\tau}\right\}^{2}\right] = \mathrm{Tr}\left[\tilde{\rho}_{\mathrm{in}}\left\{\frac{1}{\tau}\int_{0}^{\tau} dt\tilde{P}_{\mathrm{F}}(t)\right\}^{2}\right]. \tag{14.19}$$

Following steps similar to those for the thermal conductivity, we then obtain the Green-Kubo formula for $\eta$,

$$\eta = \frac{V}{k_B T}\int_{0}^{\infty} dt\tilde{C}_P(t), \tag{14.20}$$

where

$$\tilde{C}_P \equiv \mathrm{Tr}\left[\tilde{\rho}_{\mathrm{in}}\frac{1}{2}\left\{\tilde{P}_{\mathrm{F}}(t)\tilde{P}_{\mathrm{F}}(0)+\tilde{P}_{\mathrm{F}}(0)\tilde{P}_{\mathrm{F}}(t)\right\}\right]. \tag{14.21}$$

## XV. Conclusions

In this article, we have extended the Crooks-Kawasaki relation (CKR) to general quantum systems driven out of equilibrium and shown that it unifies all the results derived from the Crooks transient fluctuation theorem and the standard linear response theory derived from the Kawasaki non-linear response relation. As the CKR and the results derived from it can be readily modified for classical systems, the CKR is a unifying relation for both classical and quantum systems driven out of equilibrium.

**Acknowledgments**

I wish to thank Michele Bock for constant support and encouragement and Richard F. Martin, Jr. and other members of the physics department at Illinois State University for creating a supportive academic environment.